\documentclass[twocolumn,10pt,pra,aps,superscriptaddress,showpacs,amsmath,amssymb,floatfix]{revtex4-1}
\usepackage{graphicx} 

\usepackage{xcolor}

\usepackage{braket} 
\usepackage{bm, upgreek}
\usepackage{amsthm}
\usepackage{mathtools} 

\begin{document}

\title{A Review of Variational Quantum Algorithms: Insights into Fault-Tolerant Quantum Computing}

\author{Zhirao Wang}\thanks{These authors contributed equally to this work.}
\affiliation{Center on Frontiers of Computing Studies, Peking University, Beijing 100871, China}
\affiliation{School of Computer Science, Peking University, Beijing 100871, China}

\author{Junxiang Huang}\thanks{These authors contributed equally to this work.}
\affiliation{Center on Frontiers of Computing Studies, Peking University, Beijing 100871, China}
\affiliation{School of Computer Science, Peking University, Beijing 100871, China}

\author{Runyu Ye}\thanks{These authors contributed equally to this work.}
\affiliation{School of Science and Engineering, The Chinese University of Hong Kong, Shenzhen, Shenzhen 518172, China}
\affiliation{Shenzhen International Quantum Academy, Shenzhen 518048, China}
\affiliation{Center on Frontiers of Computing Studies, Peking University, Beijing 100871, China}

\author{Qingyu Li}
\affiliation{Center on Frontiers of Computing Studies, Peking University, Beijing 100871, China}
\affiliation{School of Computer Science, Peking University, Beijing 100871, China}

\author{Qi-Ming Ding}
\affiliation{Center on Frontiers of Computing Studies, Peking University, Beijing 100871, China}
\affiliation{School of Computer Science, Peking University, Beijing 100871, China}

\author{Yiming Huang}
\affiliation{Institute of High Energy Physics, Chinese Academy of Sciences, Beijing 100049, China}
\affiliation{China Center of Advanced Science and Technology, Beijing 100190, China }
\affiliation{Center on Frontiers of Computing Studies, Peking University, Beijing 100871, China}
\affiliation{School of Computer Science, Peking University, Beijing 100871, China}

\author{Ting Zhang}
\affiliation{Center on Frontiers of Computing Studies, Peking University, Beijing 100871, China}
\affiliation{School of Computer Science, Peking University, Beijing 100871, China}

\author{Yumeng Zeng}
\affiliation{Center on Frontiers of Computing Studies, Peking University, Beijing 100871, China}
\affiliation{School of Computer Science, Peking University, Beijing 100871, China}

\author{Jianshuo Gao}
\affiliation{Center on Frontiers of Computing Studies, Peking University, Beijing 100871, China}
\affiliation{School of Computer Science, Peking University, Beijing 100871, China}

\author{Xiao Yuan}
\email{xiaoyuan@pku.edu.cn}
\affiliation{Center on Frontiers of Computing Studies, Peking University, Beijing 100871, China}
\affiliation{School of Computer Science, Peking University, Beijing 100871, China}

\author{Yuan Yao}
\email{yuan.yao@pku.edu.cn}
\affiliation{Center on Frontiers of Computing Studies, Peking University, Beijing 100871, China}
\affiliation{School of Computer Science, Peking University, Beijing 100871, China}

\begin{abstract}
    Variational quantum algorithms (VQAs) have established themselves as a central computational paradigm in the Noisy Intermediate-Scale Quantum (NISQ) era. By coupling parameterized quantum circuits (PQCs) with classical optimization, they operate effectively under strict hardware limitations. However, as quantum architectures transition toward early fault-tolerant (EFT) and ultimate fault-tolerant (FT) regimes, the foundational principles and long-term viability of VQAs require systematic reassessment. This review offers an insightful analysis of VQAs and their progression toward the fault-tolerant regime. We deconstruct the core algorithmic framework by examining ansatz design and classical optimization strategies, including cost function formulation, gradient computation, and optimizer selection. Concurrently, we evaluate critical training bottlenecks, notably barren plateaus (BPs), alongside established mitigation strategies. The discussion then explores the EFT phase, detailing how the integration of quantum error mitigation and partial error correction can sustain algorithmic performance. Addressing the FT phase, we analyze the inherent challenges confronting current hybrid VQA models. Furthermore, we synthesize recent VQA applications across diverse domains, including many-body physics, quantum chemistry, machine learning, and mathematical optimization. Ultimately, this review outlines a theoretical roadmap for adapting quantum algorithms to future hardware generations, elucidating how variational principles can be systematically refined to maintain their relevance and efficiency within an error-corrected computational environment.
\end{abstract}

\date{\today}

\maketitle

\section{Introduction}

Quantum computing uses the principles of quantum mechanics, such as superposition and entanglement, to process information in a way that differs from classical computation. This approach offers the potential for quantum advantage and exponential speedups in solving complex problems, such as simulating many-body quantum systems and optimizing large networks. However, translating this theoretical potential into practical applications relies on the design of reliable quantum algorithms. These algorithms act as the software layer needed to utilize quantum effects and outperform classical computers.

In practice, the design of these algorithms is limited by the physical constraints of the underlying quantum hardware. Specifically, algorithmic feasibility and performance are restricted by available qubit counts, achievable circuit depths, and error rates~\cite{preskill2018quantum,cerezo2021variational,bharti2022noisy,kandala2017hardware}. Based on the development of hardware capabilities, the evolution of quantum computing can be divided into several stages: from the NISQ era~\cite{preskill2018quantum} to an early fault-tolerant (EFT) regime~\cite{katabarwa2024early}, and ultimately to fault-tolerant (FT) quantum computing~\cite{shor1995scheme,aharonov2008fault}. Each stage has different hardware characteristics and requires different algorithmic design principles and performance expectations.


In the NISQ regime, quantum devices typically consist of tens to hundreds of physical qubits and operate with non-negligible physical error rates, commonly at the $10^{-3}$ level and, in some platforms, reaching the $10^{-2}$ regime~\cite{preskill2018quantum,cerezo2021variational,bharti2022noisy,acharya2023suppressing}. Circuit depth is severely limited due to noise accumulation, rendering long coherent quantum evolutions infeasible~\cite{preskill2018quantum,cerezo2021variational,bharti2022noisy}. Algorithms designed for this regime must therefore rely on shallow circuits, tolerate significant noise, and extract useful information from noisy expectation values rather than from deep, coherent dynamics~\cite{cerezo2021variational,bharti2022noisy}. Within this context, variational quantum algorithms (VQAs) emerged as a primary algorithmic framework~\cite{peruzzo2014variational,cerezo2021variational,bharti2022noisy}. By combining parameterized quantum circuits (PQCs) with classical optimization routines, VQAs enable meaningful quantum computations under stringent hardware constraints~\cite{peruzzo2014variational,cerezo2021variational}. Their relatively shallow circuit requirements and partial robustness to certain noise sources made them particularly well suited to NISQ devices, establishing VQAs as a central paradigm for early applications in quantum chemistry, materials science, optimization, and machine learning~\cite{cerezo2021variational,bharti2022noisy,kandala2017hardware}. These theoretical advances are supported by experimental demonstrations across superconducting processors~\cite{kandala2017hardware,google2020hartree,ma2025experimental}, trapped-ion platforms~\cite{wang2024demonstration,kirmani2025variational}, and photonic architectures~\cite{peruzzo2014variational,hu2025photonic}, which collectively establish the variational paradigm within the NISQ era.

The EFT regime is expected to emerge as quantum error correction (QEC) becomes partially available yet remains resource-intensive~\cite{katabarwa2024early}. In this intermediate phase, only a limited number of logical qubits can be realized. Although logical error rates can be substantially suppressed relative to physical error rates, both circuit depth and total runtime remain constrained by the overhead of error correction~\cite{katabarwa2024early,acharya2023suppressing}. Algorithms operating in this regime cannot fully rely on large-scale fault-tolerant constructions, yet they must deliver accuracies that surpass what is realistically achievable on uncorrected NISQ hardware~\cite{katabarwa2024early}. Recent progress across multiple hardware platforms has demonstrated concrete steps toward this architecture, including the realization of small logical qubits, repeated syndrome extraction, and elementary logical operations~\cite{acharya2023suppressing,paetznick2024demonstration,bluvstein2026fault,aghaee2025scaling,postler2022faulttolerantgate}. This transition naturally shifts the research focus from algorithms designed solely to cope with noise toward approaches that can leverage partial error correction under nontrivial resource constraints~\cite{preskill2025beyond}.

The ultimate FT regime assumes the availability of a large number of logical qubits protected by QEC, with logical error rates on the order of $10^{-9}$ or lower~\cite{preskill2025beyond}. This enables arbitrarily deep circuits and supports algorithms such as quantum phase estimation and large-scale Hamiltonian simulation, albeit at the cost of substantial physical resource overhead. This hardware evolution raises a structural question: \emph{what role do VQAs play in the FT era?} While VQAs were originally developed to address the limitations of NISQ hardware, the variational paradigm itself is not inherently tied to noisy devices. Recent studies suggest that variational methods may continue to offer value within logical, error-corrected spaces, for instance through variational optimization in encoded subspaces, variational compilation of fault-tolerant circuits, and hybrid quantum-classical routines operating at the logical level~\cite{johnson2017qvector,xu2021variational,watkins2024high,meyer2025learning,sayginel2023fault}.

As the community transitions toward fault-tolerant architectures, there is a distinct need to reassess the foundations, capabilities, and long-term viability of VQAs. Rather than being rendered obsolete by the emergence of fault-tolerant quantum computers, VQAs may enter a renewed phase of relevance, provided that their frameworks are systematically adapted for this new computational landscape. To address this evolution, this review provides an integrated perspective on the trajectory of VQAs. We first summarize their theoretical underpinnings and major NISQ era achievements. We then examine the challenges that shaped their development, including barren plateaus (BPs)~\cite{mcclean2018barren}, expressivity and trainability trade-offs, and classical optimization bottlenecks. Finally, we explore how variational principles can be reformulated to maintain their utility in both EFT and ultimate FT regimes.

This review is structured as follows: \textbf{Section \ref{sec:overviewofVQAarchitecture}} decomposes the core architecture of VQAs, encompassing variational quantum ansatz design, classical optimization strategies, including cost function formulation, gradient computation, optimizer selection and system-level co-design under hardware constraints, along with key training bottlenecks and their mitigation. \textbf{Section \ref{sec:faulttolerantregime}} explores the adaptation of VQAs for fault-tolerant quantum computing, covering quantum error mitigation in the NISQ era, partial quantum error correction for EFT systems, and the challenges and prospects of VQAs in the ultimate FT regime. \textbf{Section \ref{sec:applicationofVQAs}} surveys VQA applications across quantum physics, mathematical optimization, quantum chemistry, and information science. Finally, we conclude with a summary of key advances, open challenges, and the long-term roadmap for variational principles in future fault-tolerant quantum computing.

\section{Overview of VQA Architectures}
\label{sec:overviewofVQAarchitecture}
\begin{figure*}
    \centering
    \includegraphics[width=0.8\linewidth]{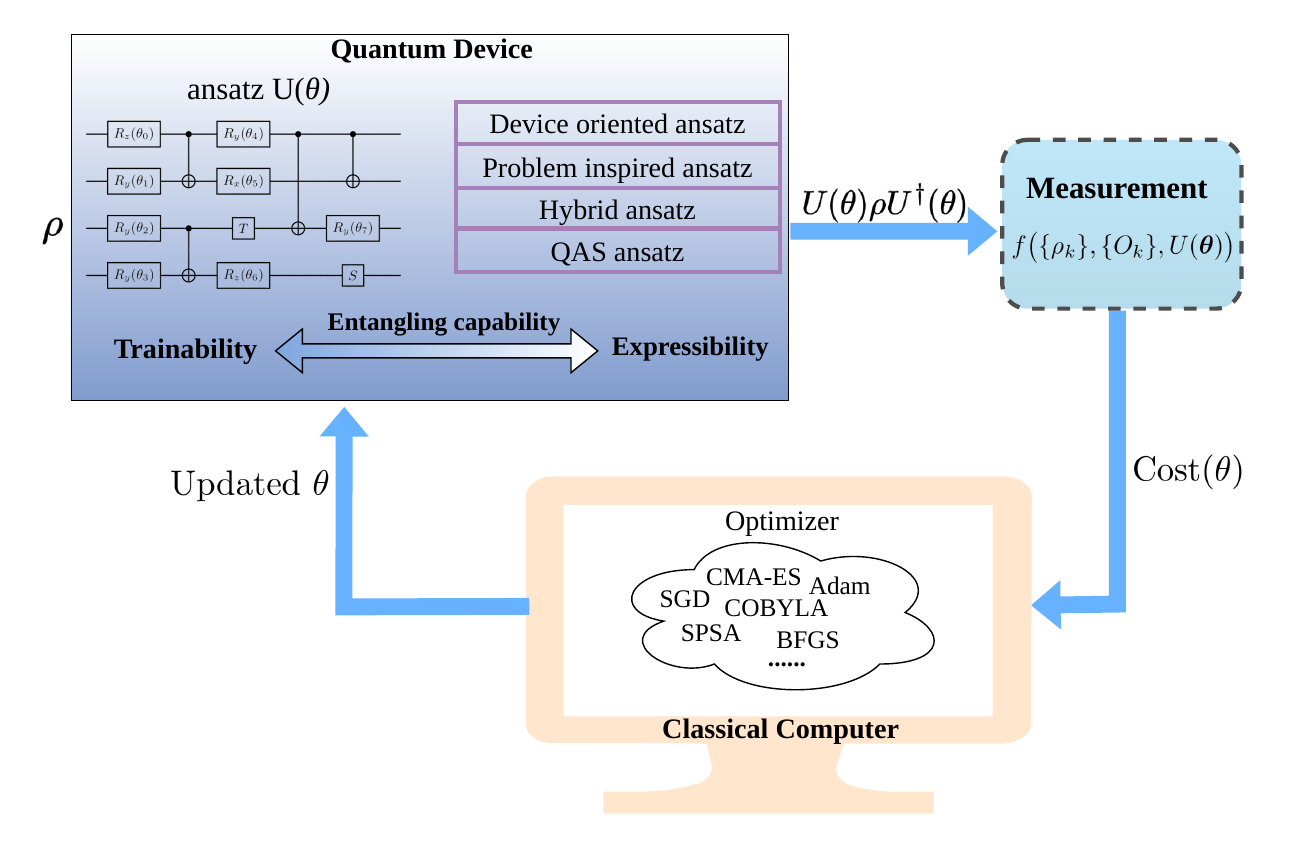}
    \caption{Architecture of a VQA: Ansatz Design, Measurement and Classical Optimization Loop.}
    \label{fig:vqaframe}
\end{figure*}

VQAs have emerged as one of the most prominent strategies for leveraging NISQ devices to achieve computational advantages. In a VQA, as illustrated in Fig.~\ref{fig:vqaframe}, the computational task is encoded into a PQC, known as the ansatz, whose parameters are collected in a vector $\bm{\theta}$, which may include continuous variables, discrete choices, or a hybrid combination depending on the application. The algorithm operates within a hybrid quantum-classical loop: the quantum device prepares input states, applies the parameterized unitary \(U(\bm{\theta})\), and measures expectation values of relevant observables to evaluate the cost function \(C(\bm{\theta})\), while a classical optimizer iteratively adjusts \(\bm{\theta}\) to minimize this function, yielding the optimal parameters
\begin{equation}
    \bm{\theta}^{\star} = \arg\min_{\bm{\theta}} C(\bm{\theta}).
\end{equation}
This paradigm integrates the expressive power of quantum state preparation and evolution with the flexibility of classical learning and optimization techniques, making VQAs a leading approach for near-term quantum advantage.

\subsection{Parameterized Quantum Circuits}
A PQC, also referred to as a variational quantum ansatz, constitutes a central component of VQAs. By specifying how the parameters $\bm{\theta}$ are embedded into the quantum circuit, the ansatz directly governs the expressive power of the model and induces the geometric structure of the resulting optimization landscape. In general, it can be represented as a product of sequential circuit layers, i.e.
$U(\bm{\theta})
    = U_l(\bm{\theta}_l)\cdots U_2(\bm{\theta}_2) U_1(\bm{\theta}_1),
$
where $\bm{\theta}_\ell$ collects the parameters associated with layer $\ell$.
Each layer can be further decomposed as
$
U_\ell(\bm{\theta}_\ell)
= \prod_m e^{-i \theta_{\ell,m} H_{\ell,m}}W_{\ell,m}.
$
where ${H_{\ell,m}}$ are Hermitian generators, the real parameters ${\theta_{\ell,m}}$ constitute the components of $\bm{\theta}_\ell$, and ${W_{\ell,m}}$ denote unparameterized quantum gates. In the following, we will discuss various extensions and generalizations of PQCs.
\subsubsection{Expressibility, entangling capability and trainability}

We first analyze three fundamental properties of variational quantum ansatzes: \textit{expressibility}, \textit{entangling capability}, and \textit{trainability}, which jointly determine their representational power and optimization efficiency on quantum hardware~\cite{sim2019expressibility,bharti2022noisy,qi2024variational}.
\textit{Expressibility} is commonly quantified by comparing the distribution of states generated by an ansatz, with randomly sampled parameters, to the uniform Haar distribution over the Hilbert space~\cite{sim2019expressibility,bharti2022noisy}. This comparison can be formalized via the Hilbert–Schmidt distance between their \(t\)-th moment operators,
\begin{equation}
    A^{(t)}_U
    = \left\|
        \int d\psi_{\mathrm{Haar}}\,
        (\lvert\psi\rangle\langle\psi\rvert)^{\otimes t}
      - \int d\bm{\theta}\,
        \bigl(\lvert\psi(\bm{\theta})\rangle
        \langle\psi(\bm{\theta})\rvert\bigr)^{\otimes t}
      \right\|_{\mathrm{HS}},
\end{equation}
where smaller $A^{(t)}_U$ indicates higher expressibility~\cite{bharti2022noisy}. However, highly expressive ansatzes may approximate unitary $t$-designs, leading to BPs in which gradients vanish exponentially with system size~\cite{mcclean2018barren,cerezo2021variational,qi2024variational}.
The \textit{entangling capability} of an ansatz is often measured by the Meyer-Wallach (Q) measure, which captures the average linear entropy of single-qubit reduced states~\cite{sim2019expressibility,bharti2022noisy}. While increasing entangling gates and connectivity enhances expressibility, it also amplifies noise and can worsen BP  effects~\cite{bharti2022noisy,qi2024variational}.
As a result, the \textit{trainability} requires a careful trade-off: shallow circuits may lack sufficient expressibility, whereas deep circuits risk flat optimization landscapes and vanishing gradients~\cite{bharti2022noisy,endo2021hybrid,qi2024variational}. To mitigate these issues, several strategies have been proposed, including (i) local cost functions to improve gradient scaling, (ii) layer-wise or block-wise training schemes, (iii) tensor-network-inspired architectures or pre-training to restrict the search space, and (iv) symmetry constraints to reduce the accessible state manifold~\cite{bharti2022noisy,qi2024variational,cerezo2021variational}.

\subsubsection{Problem-inspired ansatzes}
Problem-inspired ansatzes incorporate physical structure or prior knowledge directly into the circuit to restrict the state space to relevant manifolds, reducing circuit depth and parameter counts~\cite{bharti2022noisy,qi2024variational}. In quantum chemistry, the Unitary Coupled-Cluster Singles and Doubles (UCCSD) and truncated variational Hamiltonian ansatzes efficiently capture electron correlations by including only low-order excitations or physically significant Hamiltonian terms, while quantum information guided designs tailor entanglement using classical approximations to further enhance expressibility and shallow circuit performance~\cite{peruzzo2014variational,possel2025truncated,materia2024quantum,motta2023bridging}. For combinatorial optimization, Quantum Approximate Optimization Algorithm (QAOA) alternates problem and mixer Hamiltonians to encode task structure, effectively discretizing an adiabatic evolution~\cite{farhi2014quantum}. In quantum many-body systems, the Hamiltonian variational ansatz (HVA) leverages constituent terms of the Hamiltonian to construct layered circuits that approximate ground states with high expressibility, though compiling to native gates can require substantial depth and connectivity~\cite{wecker2015progress,qi2024variational}.

\subsubsection{Device-oriented ansatzes}

Hardware-efficient ansatzes are PQCs tailored to the native gate set and connectivity of near-term devices, achieving low depth and high implementability for variational algorithms~\cite{bharti2022noisy,cerezo2021variational,endo2021hybrid}. Formally, constrained hardware-efficient ansatzes enhance scalability and layer efficiency through principles such as universality and size consistency, though the absence of problem-specific guidance can increase parameter counts and complicate optimization~\cite{materia2024quantum}. Pulse-level variational approaches further improve performance by directly optimizing native control pulses, reducing noise, latency, and compilation overhead, and achieving chemical accuracy with significant schedule compression across superconducting, cross-resonance, and Rydberg platforms~\cite{meirom2023pansatz,liang2022pan,liang2024napa,sherbert2025parametrization,egger2023pulse,nagao2025pulse}.

\subsubsection{Hybrid ansatz and architecture-level trade-offs}
\label{sec:hybrid-ansatz}
Recent ansatz designs increasingly leverage classical resources to mitigate quantum hardware limitations, forming a hybrid paradigm between quantum and classical computation~\cite{bharti2022noisy,cerezo2021variational}. Linear combinations of parameterized quantum states optimize both classical coefficients and quantum parameters, while circuit knitting and automated architecture search decompose large-scale problems into smaller sub-circuits, reducing qubit requirements at the cost of classical post-processing~\cite{wu2025efficient}. Entanglement forging further partitions weakly entangled subsystems to effectively double hardware capacity~\cite{eddins2022doubling}. Complementary strategies use tensor networks to simulate low-entanglement layers or provide classical warm-starts, enhancing trainability, avoiding BPs, and reducing quantum circuit depth~\cite{huang2023tensor,rudolph2023synergistic}.

We emphasize that these methods \emph{change the effective problem size or circuit architecture at the ansatz level};
a complementary perspective that treats constraints at the \emph{system level} (compilation, measurement scheduling,
pulse control, and runtime latency) is discussed separately in Sec.~\ref{sec:codesign}.

\subsubsection{Quantum architecture search}

Quantum Architecture Search (QAS) automates the design of variational quantum circuits, optimizing gate types, connectivity, and depth for task performance and hardware constraints, analogous to neural architecture search in classical deep learning~\cite{pirhooshyaran2021quantum,zhang2022differentiable}. Evolutionary and heuristic methods iteratively generate and select candidate circuits via mutation, crossover, or random search, uncovering unconventional architectures at the cost of exponential search complexity~\cite{du2022quantum,li2025aqea}. Reinforcement learning approaches formulate circuit construction as a sequential decision problem, enabling policy-based optimization of compact, hardware-compatible architectures with improved sample efficiency and support for multi-objective rewards~\cite{sogabe2022model,kuo2021quantum,rapp2025reinforcement}. Differentiable QAS further relaxes discrete circuit parameters into continuous variables, allowing end-to-end gradient-based optimization of both circuit parameters and topology, supporting hardware-aware co-design, though it requires careful regularization to avoid bias from smooth relaxations~\cite{zhang2022differentiable,wu2023quantumdarts,he2024gradient}.

\subsection{Optimization}

Optimization in VQAs aims to adjust the parameters of the ansatz toward values that minimize the cost function. Various classical optimizers are used during training. In this subsection, we introduce cost functions, gradients, optimizers, and measurement-related techniques relevant to optimization.

\subsubsection{Cost function}

The cost function \(C(\boldsymbol{\theta})\) plays a central role in VQAs, as it guides the classical optimization process toward an approximate solution. Its design is therefore crucial: it must encode the objective of the problem, remain efficiently evaluable on quantum hardware, and induce an optimization landscape that is sufficiently trainable. In its most general form, the cost function defines a mapping from the trainable parameters \(\boldsymbol{\theta}\) to the real numbers,
\begin{equation}
    C(\boldsymbol{\theta}) = f\bigl(\{\rho_k\}, \{O_k\}, U(\boldsymbol{\theta})\bigr),
\end{equation}
where \(f\) is a classical post-processing function, \(U(\boldsymbol{\theta})\) is a parameterized unitary, \(\{\rho_k\}\) denotes a set of input states, and \(\{O_k\}\) is a collection of observables. This expression is often written more explicitly as
\begin{equation}
    C(\boldsymbol{\theta}) =
    \sum_k f_k\!\left(\operatorname{Tr}\!\big[O_k\, U(\boldsymbol{\theta})\,
    \rho_k\, U^{\dagger}(\boldsymbol{\theta})\big]\right),
\end{equation}
with \(\{f_k\}\) being scalar-valued functions.

In many practically relevant VQAs, especially variational eigensolvers, the cost function reduces to the expectation value of a Hamiltonian \(H\) with respect to the parameterized quantum state \(|\psi(\boldsymbol{\theta})\rangle = U(\boldsymbol{\theta})|0\rangle\), namely
\begin{equation}
    C(\boldsymbol{\theta}) = \langle \psi(\boldsymbol{\theta}) | H | \psi(\boldsymbol{\theta}) \rangle.
\end{equation}
Since such expectation values are not directly accessible on quantum hardware, they must be estimated empirically through repeated circuit executions and projective measurements. Typically, the Hamiltonian is decomposed into a weighted sum of Pauli strings,
\begin{equation}
    H = \sum_j c_j P_j,
\end{equation}
so that the cost can be approximated as
\begin{equation}
    C(\boldsymbol{\theta}) \approx \sum_{j=1}^m c_j \, \langle P_j \rangle_{\boldsymbol{\theta}},
\end{equation}
where each Pauli expectation value \(\langle P_j \rangle_{\boldsymbol{\theta}}\) is estimated from a finite number of measurement shots~\cite{mcclean2016theory,kandala2017hardware}. In practice, the required number of shots depends on the target precision, circuit depth, and hardware noise level.

The evaluation strategy depends on the structure of the observable \(O_k\)~\cite{qi2024variational}. When \(O_k\) is a Pauli operator, direct measurement in the appropriate basis is sufficient. More generally, Hamiltonians such as those arising in Ising-type models can often be expressed as sums of local Pauli terms, \(O_k = \sum_{j} c_{j,k} P_{j,k}\), where each \(P_{j,k}\) acts nontrivially on at most \(m\) qubits. This yields an \(m\)-local cost function, which is often more favorable for trainability and can mitigate BPs compared with global cost functions~\cite{cerezo2021cost}. For overlap-based objectives involving density operators, such as \(\operatorname{Tr}[\rho \sigma]\), dedicated protocols including the swap test~\cite{buhrman2001quantum} and destructive swap test~\cite{garcia2013swap} can be employed. Observables containing nontrivial unitary operators generally require additional basis transformations or auxiliary circuits before measurement.

Because repeated sampling can become a major bottleneck, several techniques have been developed to reduce the measurement overhead and improve estimator accuracy. Measurement grouping exploits the commutativity of Pauli operators so that multiple terms can be estimated simultaneously within a single measurement setting~\cite{verteletskyi2020measurement}. Classical shadow methods further improve sample efficiency by using randomized measurements and classical post-processing to predict many observables from a limited set of outcomes~\cite{huang2020predicting}. In addition, error mitigation techniques, such as zero-noise extrapolation, probabilistic error cancellation, readout calibration, and hardware-efficient mitigation schemes, aim to suppress the influence of hardware noise on the estimated cost values without requiring ultimate fault tolerance~\cite{temme2016error,kandala2019error,cai2023quantum,suzuki2022quantum,escudero2023hardware}. Although these methods significantly improve the practicality of near-term VQAs, accurate and efficient cost-function evaluation remains one of the major challenges in the field.

Overall, a well-designed cost function should satisfy three key requirements: its global minimum should correspond exactly to the desired solution, its expectation values should be experimentally accessible on quantum hardware, and the resulting optimization landscape should remain sufficiently well-behaved to enable efficient classical optimization.

\subsubsection{Gradient}

Gradients characterize how the cost function changes with respect to variations in the circuit parameters. Consequently, many widely used optimizers rely on gradient information to iteratively update parameters and drive convergence toward high-quality solutions.

Given a parameter vector \(\bm{\theta}\) and a cost function \(C(\bm{\theta})\), a standard gradient-descent update in VQAs is given by
\begin{equation}
    \bm{\theta}^{(t+1)}
    = \bm{\theta}^{(t)}
      - \eta^{(t)} \, \nabla_{\bm{\theta}} C\bigl(\bm{\theta}^{(t)}\bigr),
    \label{eq:gd_update}
\end{equation}
where \(\eta^{(t)}\) denotes the learning rate, which controls the trade-off between exploration and convergence and is often chosen with the constraints of NISQ devices in mind~\cite{bharti2022noisy}.

In practice, gradients can be estimated using finite-difference methods, analytic parameter-shift formulas, or other derivative rules tailored to parameterized quantum circuits~\cite{endo2021hybrid}.

\paragraph{Parameter-shift rule.}

The parameter-shift rule provides an analytic method for evaluating gradients by measuring the circuit at appropriately shifted parameter values. This approach exploits the algebraic structure of gates generated by Pauli-like operators and is applicable to a broad class of PQCs.

Specifically, if a parameter $\theta_k$ appears only in a gate of the form
\[
U_k(\theta_k) = e^{-i \theta_k G_k},
\]
where the generator $G_k$ has eigenvalues $\pm r_k/2$, then the derivative of the cost function with respect to $\theta_k$ can be written as
\begin{equation}
    \frac{\partial C(\boldsymbol{\theta})}{\partial \theta_k}
    =
    \frac{r_k}{2}
    \left[
        C\!\left(\boldsymbol{\theta} + s_k \boldsymbol{e}_k\right)
        -
        C\!\left(\boldsymbol{\theta} - s_k \boldsymbol{e}_k\right)
    \right],
\end{equation}
where $s_k = \frac{\pi}{2 r_k}$ and $\boldsymbol{e}_k$ is the unit vector in the $k$-th parameter direction. Thus, each gradient component can be estimated using two additional circuit evaluations at shifted parameter values. More general generator spectra can be treated using generalized shift rules or, when necessary, finite-difference approximations~\cite{cerezo2021variational}.

\paragraph{Higher-order derivatives.}

The parameter-shift technique can also be extended to evaluate higher-order derivatives~\cite{cerezo2021variational}. For example, applying the shift rule twice yields the second derivative:
\begin{equation}
\begin{split}
\frac{\partial^2 C}{\partial \theta_l^2} = \sum_k &\frac{1}{4\sin^2\alpha} \Big( \operatorname{Tr} \left[ O_k U(\boldsymbol{\theta} + 2\alpha \boldsymbol{e}_l) \rho_k U^\dagger(\boldsymbol{\theta} + 2\alpha \boldsymbol{e}_l) \right] \\
&+ \operatorname{Tr} \left[ O_k U(\boldsymbol{\theta} - 2\alpha \boldsymbol{e}_l) \rho_k U^\dagger(\boldsymbol{\theta} - 2\alpha \boldsymbol{e}_l) \right] \\
&- 2\operatorname{Tr} \left[ O_k U(\boldsymbol{\theta}) \rho_k U^\dagger(\boldsymbol{\theta}) \right] \Big).
\end{split}
\end{equation}
Higher-order derivatives can be obtained analogously through repeated application of the same principle~\cite{mari2021estimating}.

\paragraph{Quantum natural gradient.}

Beyond estimating derivatives in the standard Euclidean parameter space, one can also exploit the underlying geometry of the variational state space to design more effective update rules. Quantum natural gradient (QNG) extends classical steepest-descent methods to optimization on a Riemannian manifold by replacing the Euclidean metric in parameter space with the Fubini--Study metric tensor \(G(\boldsymbol{\theta})\)~\cite{amari1998natural,stokes2020quantum,yao2022complex}. In this way, QNG accounts for the geometry of quantum states and often provides a more physically meaningful search direction than conventional gradient descent, which can improve convergence in ill-conditioned optimization landscapes.

\subsubsection{Optimizers}

Optimizers iteratively update a parameterized quantum circuit's parameters \(\boldsymbol{\theta}\) to minimize a cost function \(C(\boldsymbol{\theta})\). In quantum computing, optimization faces two key challenges: stochastic cost estimates due to finite measurements and BPs which means regions with exponentially vanishing gradients~\cite{mcclean2018barren}. Optimizers are broadly split into gradient-based and gradient-free methods, each offering different trade-offs in sample complexity, noise robustness, and convergence speed. This section reviews those used in VQAs, outlining their principles and characteristics.

\paragraph{Gradient-based methods.} Gradient-based methods estimate gradients via the parameter-shift rule~\cite{schuld2019evaluating} or finite differences, then update parameters along the steepest descent.

Stochastic gradient descent (SGD) updates the parameters according to
\begin{equation}
\boldsymbol{\theta}^{(t+1)} = \boldsymbol{\theta}^{(t)} - \eta \, \nabla\hat{C}(\boldsymbol{\theta}^{(t)}),   
\end{equation}
where $\eta$ is the learning rate and $\nabla\hat{C}(\boldsymbol{\theta}^{(t)})$ denotes the estimated gradient at iteration $t$. In noisy settings, however, gradient estimates can lead to unstable or erratic convergence. To mitigate this issue, momentum introduces a velocity vector that accumulates past gradient information:
\begin{equation}
    \mathbf{u}^{(t+1)} = \beta \mathbf{u}^{(t)} + \eta \, \nabla\hat{C}(\boldsymbol{\theta}^{(t)}),
    \quad
    \boldsymbol{\theta}^{(t+1)} = \boldsymbol{\theta}^{(t)} - \mathbf{u}^{(t+1)},
\end{equation}
where $\beta \in [0,1)$ is the momentum coefficient and $\mathbf{u}^{(t)}$ denotes the velocity term.

Adam is a first-order optimization method that combines momentum and adaptive learning rates through exponentially decaying averages of past gradients (first moments) and squared gradients (second moments). With bias-corrected moment estimates, the parameter update rule is given by
\begin{equation}
    \boldsymbol{\theta}^{(t+1)} = \boldsymbol{\theta}^{(t)} - \eta \frac{\hat{\mathbf{m}}^{(t+1)}}{\sqrt{\hat{\mathbf{v}}^{(t+1)}}+\epsilon},
\end{equation}
where $\hat{\mathbf{m}}^{(t+1)}$ and $\hat{\mathbf{v}}^{(t+1)}$ denote the bias-corrected first- and second-moment estimates, respectively, and $\epsilon$ is a small positive constant added for numerical stability~\cite{kingma2014adam}.

Broyden–Fletcher–Goldfarb–Shanno (BFGS) is a quasi-Newton method that builds an approximation of the inverse Hessian matrix using a limited history of gradient updates~\cite{liu1989limited}. This approximation enables superlinear convergence without explicitly computing second-order derivatives, which is why it is often considered the gold standard for noiseless deterministic optimization problems. However, when noise is introduced, BFGS suffers degradation in performance. This makes BFGS primarily suitable for noiseless simulations or as a baseline for comparison, rather than for direct hardware implementation.

Riemannian gradient descent methods exploit the quantum state manifold's geometry. Quantum Natural Gradient (QNG), as a prominent instance, uses the Fubini–Study metric to adjust updates according to local curvature~\cite{stokes2020quantum, yao2024riemannian}. However, estimating the metric tensor costs \(\mathcal{O}(m^2)\) measurements per iteration for \(m\) parameters~\cite{jones2025benchmarking}, which can offset its practical advantages.

\paragraph{Gradient-free methods.} Given the difficulty and cost of estimating gradients on quantum hardware, gradient-free methods have become a popular and more robust choice. These methods treat the quantum circuit as a black-box function and do not require explicit gradient calculations.

Simultaneous Perturbation Stochastic Approximation(SPSA) approximates gradients with two cost evaluations per iteration, perturbing all parameters simultaneously~\cite{spall1992multivariate}:
\begin{equation}
    \nabla \hat{C}(\boldsymbol{\theta}^{(t)}) \approx \frac{
        C\!\left(\boldsymbol{\theta}^{(t)} + c_t \boldsymbol{\Delta}^{(t)}\right)
        -
        C\!\left(\boldsymbol{\theta}^{(t)} - c_t \boldsymbol{\Delta}^{(t)}\right)
    }{2c_t}
    \left(\boldsymbol{\Delta}^{(t)}\right)^{-1},
\end{equation}
where $\boldsymbol{\Delta}^{(t)}$ is a random perturbation vector, typically with independent entries taking values $\pm 1$, and $\left(\boldsymbol{\Delta}^{(t)}\right)^{-1}$ denotes the element-wise inverse.
Wiedmann et al. shows hybrid gradient-SPSA variants outperform both standard SPSA and parameter-shift methods under shot and hardware noise~\cite{wiedmann2023empirical}.

Constrained Optimization BY Linear Approximations (COBYLA) builds linear approximations within a trust region to propose new points~\cite{powell1998direct}. At each iteration, it builds a linear interpolation model using the current set of points and solves a trust-region subproblem to determine the next candidate point. It is popular for its simplicity and robustness, though tuned CMA-ES or SPSA may yield better outcomes~\cite{bonet2023performance}.

Covariance Matrix Adaptation Evolution Strategy (CMA-ES) is an evolutionary algorithm that generates new candidate solutions by sampling from a multivariate normal distribution~\cite{hansen2006cma}. It adapts the covariance matrix of the distribution to guide the search towards promising areas of the landscape, effectively learning the underlying correlation structure of good solutions. 

Nelder-Mead, also known as the simplex method, maintains a simplex of \(n+1\) points and uses reflection, expansion, and contraction to seek minima~\cite{nelder1965simplex}. While Nelder-Mead is simple and was used in some early VQE demonstrations as a baseline, it is generally slow and unreliable in higher dimensions, and modern VQA benchmarks typically favor more sophisticated optimizers like those mentioned above.

Beyond the well-established methods, recent optimizers explicitly incorporate problem structure, noise models, or generative learning. Variational optimization for quantum problems using deep generative networks (VGON)~\cite{zhang2025variational} learns a distribution over high-performing parameters via a deep generative model. Physics-informed Model for Accelerating Large-scale Quantum Optimization (PALQO)~\cite{huang2025palqo} uses physics-informed neural networks to predict multiple future updates, drastically cutting quantum evaluations. These, along with Bayesian optimization and evolutionary strategies~\cite{bonet2023performance}, represent a trend toward problem-aware optimizers maximizing performance under quantum resource constraints.

\subsection{System-level co-design under hardware limitations}\label{sec:codesign}

Beyond ansatz design (Sec.~\ref{sec:hybrid-ansatz}), the practical performance of VQAs is often dominated by
\emph{system-level} constraints: compilation overhead to native or logical gate sets, measurement scheduling and estimator
design, control latency, and (on some platforms) pulse-level feasibility. In this section we summarize a unified co-design
view in which the quantum program, hardware control stack, and classical runtime are optimized jointly.

A first co-design axis is \emph{compilation- and overhead-aware objectives}. On realistic devices, the implemented circuit
differs from the abstract ansatz due to routing, calibration constraints, and (in the fault-tolerant setting) logical gate
synthesis. A system-level strategy therefore tracks resource proxies (e.g., depth, two-qubit-gate count, or logical
non-Clifford cost) alongside the task loss, and explicitly optimizes objectives that penalize compilation overhead.
This perspective is closely related to variational compilation and resource-aware simulation routines, where the goal is to
approximate a target unitary or dynamics with bounded overhead under hardware constraints~\cite{khatri2019quantum,sharma2020noise,mizuta2022local,sanders2020compilation,watkins2024high,sayginel2023fault,dangwal2025variational,katabarwa2024early}.
In the FT regime, where gate discreteness and non-Clifford resources dominate, such overhead-aware optimization turns
variationality into a \emph{resource-aware} routine rather than a purely physics-driven search.

A second axis is \emph{measurement scheduling and estimator co-design}. Even when circuit depth is modest, the measurement
wall can dominate wall-clock runtime, so the estimator must be treated as part of the architecture. Common ingredients
include partitioning observables into commuting families~\cite{verteletskyi2020measurement,yen2020measuring},
near-optimal measurement scheduling and partial-tomography strategies~\cite{bonet2020nearly},
and modern estimators such as classical shadows and derandomization that reduce sample complexity for many targets~\cite{huang2020predicting,huang2021efficient,huggins2022nearly}.
Hardware-efficient \emph{entangled} measurement constructions further enlarge the design space of estimators and can reduce
sampling overhead for certain VQA objectives~\cite{escudero2023hardware}.

A third axis is \emph{pulse-level and control co-design}. For platforms that expose pulse shaping, variational parameters can
be moved from gate angles to native control degrees of freedom, reducing compilation latency and sometimes shortening
schedules. Pulse-level variational ansatzes and intermediate-level native-pulse compilation illustrate how co-design trades
algorithmic abstraction for hardware efficiency~\cite{meirom2023pansatz,liang2022pan,liang2024napa,egger2023pulse,sherbert2025parametrization,nagao2025pulse}.
These methods interact nontrivially with gradient estimation and optimizer choice; for instance, analytic or hardware-friendly
derivative estimators can change the end-to-end runtime profile~\cite{mari2021estimating,bittel2022fast}.

A final axis concerns \emph{runtime latency and classical--quantum feedback}. Even at a fixed shot budget, throughput depends
on queueing, reset time, classical post-processing, and outer-loop synchronization. Practical co-design includes batching
circuit evaluations, reducing parameter-update frequency (e.g., reusing measurement data when statistically valid), and
choosing latency-aware optimizers. These considerations become increasingly important as devices scale and distributed
control stacks introduce non-negligible overhead.

In EFT regimes, co-design must also include \emph{interfaces to error mitigation and partial QEC}. Here, performance is
shaped by the interaction between variational updates and mitigation/partial correction choices. A co-design view treats
mitigation not as a post-hoc patch but as part of the system architecture, aiming to minimize \emph{total} cost
(shots $\times$ depth $\times$ classical overhead) for a target accuracy~\cite{endo2021hybrid,cai2023quantum,suzuki2022quantum,temme2016error,kandala2019error,huggins2021virtual,van2023probabilistic,quek2024exponentially,strikis2021learning,ravi2022vaqem}.

Techniques that primarily \emph{increase the effective qubit capacity} by decomposing the task into subcircuits
(e.g., circuit knitting), partitioning weakly entangled subsystems (entanglement forging), or using tensor-network warm-starts
belong to ansatz and architecture design, and are covered in Sec.~\ref{sec:hybrid-ansatz}~\cite{wu2025efficient,eddins2022doubling,huang2023tensor,rudolph2023synergistic}.
Here we focus on system-level constraints that remain even after the ansatz class is fixed.

\subsection{Challenges and Potential Solutions}

VQAs are widely regarded as primary candidates for demonstrating quantum advantage on NISQ devices~\cite{cerezo2021variational, bharti2022noisy}. However, scaling these algorithms to classically intractable regimes presents fundamental challenges, primarily revolving around the ability to navigate high-dimensional optimization landscapes and the feasibility of estimating observables with sufficient precision.

A central barrier to trainability is the BPs phenomenon, where gradients vanish exponentially with system size, rendering the optimization landscape effectively flat for any polynomial number of samples~\cite{mcclean2018barren}. Extensive research has linked the origins of BPs to the mathematical properties of the cost function and the ansatz structure. While global cost functions inevitably lead to BPs, local cost functions can ensure trainability in shallow circuits~\cite{cerezo2021cost}, a property notably exploited in Quantum Convolutional Neural Networks~\cite{pesah2021absence}. Nevertheless, structural design involves a critical trade-off: the excessive ansatz expressibility directly correlates with gradient vanishing~\cite{holmes2022connecting}, and volume-law entanglement can similarly induce BPs~\cite{ortiz2021entanglement}. These insights explain why generic Hardware-Efficient Ansatzes have largely been superseded by more structured designs. Additionally, hardware noise acts as a physical source of untrainability; noise-induced BPs occur when noise drives the quantum state toward the maximally mixed state, causing the landscape to collapse as circuit depth increases~\cite{wang2021noise}.

Recent theoretical advances have unified these observations under the framework of Dynamical Lie Algebras (DLAs). This theory posits that systems with exponentially scaling DLA dimensions inherently exhibit BPs, whereas those restricted to polynomially scaling DLAs remain generally trainable~\cite{larocca2022diagnosing}. It also enables rigorous analyses of overparameterization~\cite{larocca2023theory} and landscape characterization~\cite{fontana2023adjoint}. Nevertheless, exceptions exist: even in otherwise favorable algebras, particular choices of initial states can still induce BPs~\cite{diaz2023showcasing}.
Viewed through this lens, many successful mitigation strategies can be interpreted as {constraining the effective DLA explored during training} and/or {keeping optimization within well-conditioned neighborhoods}. For example, Geometric Quantum Machine Learning incorporates physical symmetries (equivariance) directly into the ansatz, thereby restricting the accessible algebra and improving trainability~\cite{ragone2022representation, nguyen2024theory, schatzki2024theoretical}. Scalability can be further enhanced by replacing static circuits with adaptive constructions that control expressivity growth, from parameter-efficient ADAPT-type schemes~\cite{grimsley2019adaptive} to depth-efficient variants such as TETRIS-ADAPT-VQE, which enforces commutativity to yield denser circuit growth~\cite{anastasiou2024tetris}. Complementarily, deterministic initialization and warm-start methods aim to start (and remain) near benign regions of the landscape, including Identity Block Initialization~\cite{grant2019initialization, zhang2022escaping} and classical- or tensor-network-assisted warm starts~\cite{egger2021warm, rudolph2023synergistic}. Additional algorithmic tools include analyzing parameter distributions via ZX-calculus~\cite{zhao2021analyzing} and layerwise training protocols~\cite{skolik2021layerwise}. 


However, achieving trainability may come at a cost. 
The trainability-simulatability dilemma suggests that constraints that stabilize gradients (e.g., restricting dynamics to small Lie algebras) may also render circuits efficiently simulatable by classical methods, potentially reducing the prospect of quantum advantage~\cite{cerezo2025does}.
Consequently, dynamics that are hard to simulate, such as unitary scramblers, are intrinsically hard to train~\cite{holmes2021barren}. Furthermore, even when BPs are avoided, landscapes may still suffer from ill-conditioned Hessian matrices (narrow gorges)~\cite{cerezo2021higher}, which suppress both gradient-based and gradient-free optimization~\cite{arrasmith2021effect}.

Beyond finding optimal parameters, the efficiency of measuring the cost function poses another significant bottleneck. The number of shots required for precision $\epsilon$ scales as $O(1/\epsilon^2)$, making naive measurements of complex Hamiltonians prohibitively expensive. Initial solutions utilized operator grouping to measure commuting families simultaneously~\cite{gokhale2019minimizing, yen2020measuring}, with specialized scheduling for fermionic systems~\cite{bonet2020nearly}. A major advancement was the introduction of Classical Shadows, which allow for the prediction of numerous observables with logarithmic sample complexity~\cite{huang2020predicting}. This technique has been further refined through quantum overlapping tomography~\cite{cotler2020quantum}, derandomized shadows to minimize variance~\cite{huang2021efficient}, and shadow distillation for error mitigation~\cite{seif2023shadow}.

As the field moves toward EFT hardware, both trainability and efficiency are shaped by new constraints. Logical gate sets can be effectively discrete (e.g., Clifford+T), and high-fidelity, error-corrected measurements may carry substantial time costs. This motivates the notion of a measurement wall, where straightforward extensions of shot-intensive VQA workflows become temporally impractical due to the runtime overhead of repeated logical measurements~\cite{gonthier2022measurements}.
This necessitates a shift toward sample-efficient strategies compatible with discrete landscapes, such as Bayesian optimization~\cite{sorourifar2024bayesian}, and algorithms operating near the theoretical limits of information extraction~\cite{huggins2022nearly, o2022quantum}. Architecture-aware designs, such as EFT-VQAs with Partial Error Correction, are also emerging to manage logical operation costs by balancing accuracy against runtime~\cite{dangwal2025variational}.

\section{Fault-tolerant Regime of VQAs}
\label{sec:faulttolerantregime}
VQAs have become a promising paradigm in near-term quantum computing, benefiting from a highly flexible framework that leverages carefully designed ansatzes and well-defined loss functions together with classical optimizers.
This enables them to adaptively tune the parameters of PQCs according to device characteristics, thereby exhibiting a certain degree of resilience to quantum noise.
However, their hybrid nature also makes them highly sensitive to decoherence and gate noise: small stochastic errors can distort the cost landscape, slow convergence, and obscure the true energy minimum. 
As quantum hardware advances toward the era of early fault tolerance, the integration of Quantum Error Mitigation (QEM) and QEC techniques with VQAs has emerged as one of the most promising strategies for designing more robust and efficient VQAs, with the potential to be applied to practically valuable tasks, such as quantum simulation and quantum optimization, and even to demonstrate quantum advantage.

Broadly speaking, the integration of error resilience into VQAs can be viewed as a continuum that parallels the technological evolution of quantum hardware:
\begin{itemize}
    \item In the NISQ regime, VQAs are executed directly on physical qubits and rely primarily on error mitigation techniques to reduce the impact of quantum noise. At this stage, VQAs are mainly used for proof of concept;
    \item In the EFT regime~\cite{katabarwa2024early}, QECs are incorporated into the design of VQAs to encode logical qubits and protect noise-sensitive operations, while error mitigation is still employed to further suppress the effects of noise on the remaining parts of the computation. In this stage, VQAs may begin to demonstrate practical utility in selected settings.
    \item In the FT regime, VQAs operate entirely on logical qubits, and all quantum operations are protected by QEC. At this stage, VQAs could become useful for selected applications across a broad range of domains.
\end{itemize}
This progression from NISQ to EFT to ultimate FT delineates the roadmap by which VQAs are expected to evolve toward scalable, noise-resilient computation.

\subsection{NISQ era}
Since the advent of NISQ devices, substantial effort has gone into understanding what VQAs can achieve under realistic hardware constraints. This body of work positions VQAs not only as heuristics for specific tasks, but also as a general framework for benchmarking and probing the capabilities of noisy, shallow quantum circuits.
However, under current hardware limitations, many low-depth noisy VQAs remain efficiently classically simulable in practically relevant regimes~\cite{shao2024simulating,schuster2025polynomial,fontana2025classical}, making it challenging to demonstrate a clear quantum advantage. Moreover, many existing demonstrations remain proof-of-principle studies rather than immediately useful applications; they are more like proofs of concept than tools with immediate utility.
Nevertheless, studying VQAs in the NISQ setting remains important: it helps clarify the operational boundaries of near-term quantum devices and provides guidance for designing more noise-resilient algorithms and quantum hardware in the early fault-tolerant regime. It can further advance VQAs toward practical utility, ultimately paving the way for demonstrating quantum advantage in the era of fault-tolerant quantum computation.
Within this landscape, this section reviews recent progress on the integration of VQAs with QEM and QEC techniques on near-term quantum hardware with enhanced error resilience, focusing on quantum error mitigation, quantum variational error corrector~(QVECTOR), and the design of QEC codes by VQAs.

\subsubsection{VQA-based Error Mitigation}
Before we enter the era of fault-tolerant quantum computing, the impact of quantum noise on quantum computation remains an unavoidable and formidable challenge. To confront this challenge in the NISQ era, quantum error mitigation has been proposed as a transitional approach. Rather than eliminating errors during computation, it reduces the adverse effects of noise through a collection of mathematical and physical ideas, such as statistical inference, equivalence transformations and symmetry conservation, thereby improving the reliability of quantum algorithms.
In the development of this field, foundational techniques such as zero-noise extrapolation and probabilistic error cancellation have been established as the baseline for noisy computation~\cite{temme2016error, kandala2019error}.
The field has subsequently expanded to encompass data-driven approaches, exemplified by Clifford data regression~\cite{czarnik2021error}, and physics-inspired methods like virtual distillation, which utilizes state interference to suppress incoherent errors~\cite{huggins2021virtual}. 
A few comprehensive overviews of these mitigation techniques are provided in~\cite{tilly2022variational,cai2023quantum}.
The maturity of this domain was recently underscored by the demonstration that advanced probabilistic error cancellation, utilizing sparse Pauli-Lindblad noise models~\cite{van2023probabilistic}, enables accurate estimation on large-scale processors beyond the reach of brute-force classical simulation~\cite{kim2023evidence}.

While the VQA paradigm inherently possesses some resilience against coherent errors such as over-rotations~\cite{sharma2020noise}, incoherent noise constitutes a critical bottleneck that directly degrades expectation values. 
Beyond applying post-processing error-mitigation strategies such as zero-noise extrapolation and probabilistic error cancellation~\cite{endo2021hybrid} to improve the performance of VQAs, another meaningful research direction is to investigate whether variational techniques themselves can be incorporated into an error-mitigation framework to suppress noise.

Following this line of research, several works have proposed VQA-based error-mitigation methods that embed the mitigation procedure directly into the variational optimization loop. For example, the Variational Approach to Quantum Error Mitigation framework~\cite{ravi2022vaqem} formulates error mitigation as a variational learning task by optimizing circuit parameters to compensate for hardware noise.
Learning-based approaches~\cite{strikis2021learning} train classical models on noisy measurement data to infer noise-mitigated observables, which can be integrated into standard VQA workflows.
Some studies have further investigated whether variational noise-mitigation strategies can improve the trainability of VQAs under realistic noise models~\cite{wang2024can}.
More specialized works, such as variational noise mitigation for structured circuits like the quantum Fourier transform~\cite{gomez2025variational}, demonstrate that variational mitigation can also be tailored to algorithm-specific settings.
Although these approaches are promising for enhancing VQA performance on near-term devices, they do not yet provide the robust and scalable protection required for fault-tolerant quantum computation.
Moreover, the applicability of QEM is constrained by sampling overheads that grow exponentially with circuit depth~\cite{quek2024exponentially}, which also poses a major challenge and may prevent VQAs from demonstrating a quantum advantage in the NISQ era.

\subsubsection{Quantum variational error corrector algorithm}
Beyond combining VQAs with error mitigation, integrating them with ideas from quantum error correction theory offers an alternative approach.
Quantum error-correcting codes such as the color code and the surface code require a large number of physical qubits to encode a single logical qubit for fault-tolerant quantum computation, making their deployment on current quantum computers, which typically have only hundreds of qubits, impractical.
QVECTOR provides an end-to-end, device-tailored quantum--classical hybrid approach to protecting quantum information using variational quantum circuits. 
Inspired by the general paradigm of quantum error correction, it structures the procedure into three stages: encoding, recovery and decoding. 
In the quantum memory setting, QVECTOR shows strong potential; under phase-damping noise, simulations indicate that it can extend the effective $T_2$ time of a quantum memory by roughly a factor of six~\cite{johnson2017qvector}.

In QVECTOR, the system is partitioned into logical, syndrome, and refresh qubits. A logical input state is first encoded by a parameterized circuit \(V(\boldsymbol{\theta}_1)\) into the logical-syndrome space, after which a recovery circuit \(W(\boldsymbol{\theta}_2)\) acts on the full system. The refresh qubits are then discarded, inducing an effective non-unitary operation that helps suppress decoherence. Finally, the inverse encoding \(V^\dagger(\boldsymbol{\theta}_1)\) is applied to recover the original logical state.

The parameters of \(V\) and \(W\) are trained by maximizing an average fidelity-based objective over input states drawn from a unitary \(2\)-design, improving generalization beyond a small training set. A key advantage of QVECTOR is that it does not rely on an explicit noise model; instead, it learns encoding and recovery directly from data sampled on the noisy device. This makes it a practical, device-tailored method for near-term quantum memory protection, although it does not provide the rigorous scalability guarantees of fault-tolerant QEC~\cite{johnson2017qvector}.

\subsubsection{Variational Design of QEC Codes}
As a cornerstone of quantum error correction theory, QECCs are of great significance. Along the development of quantum error correction theory, various constructions of QECCs have been developed. However, the majority of these codes are designed based on general quantum noise models, which often fail to accurately capture the intricate noise characteristics of real quantum devices, thereby limiting their practical performance and making them unsuitable for near-term quantum devices.
To address this challenge, several recent works, including the Variational Circuit Compiler (VCC)~\cite{xu2021variational}, Variational Quantum Error Correction (VarQEC)~\cite{cao2022quantum}, and Variational Graphical Quantum Error Correction (VGQEC)~\cite{shao2024variational}, have been proposed for the design of device-tailored quantum error-correcting codes for logical qubits, aiming to obtain optimal hardware-adapted quantum code constructions.

Instead of utilizing ad-hoc or analytically derived circuits in VQAs, VCC~\cite{xu2021variational} maps the problem of preparing logical code states into a variational optimization task that minimizes noisy operations and adapts to native gate sets and qubit connectivity. They demonstrate this approach by deriving optimized encoders for the five-qubit and seven-qubit codes, showing that the variational compiler can produce shallower, hardware-friendly encoding circuits that reduce error accumulation on NISQ devices.

VGQEC method~\cite{shao2024variational} introduces a novel approach, namely the Quon 3D graphical framework, for extending various QEC codes tailored to specific noise environments based on established general-purpose codes.
By inserting parameterized cross-points into the graphical representation of known codes, like repetition or stabilizer codes, VGQEC dynamically reconfigures the code structure by adjusting these parameterized nodes to optimize for specific noise environments, achieving higher performance than the original generic code.
This approach leverages prior knowledge to ensure faster, more stable optimization and interpretability compared to designing from scratch, even allowing smooth transitions between different code types (e.g., bridging repetition and $[[5,1,3]]$ codes). 
The numerical simulations show that, compared with QVECTOR, VGQEC provides superior error correction performance with fewer parameters and optimization steps.
However, this reliance on existing structures can also be a limitation, potentially restricting its ability to discover radically novel codes. 

VarQEC~\cite{cao2022quantum} further generalizes the use of VQAs from compiling known QECCs to discovering QECCs themselves by formulating QEC as a learning problem guided by cost functions derived from the Knill–Laflamme conditions. 
Rather than starting from a predefined stabilizer structure, VarQEC uses a PQC to jointly optimize the code’s basis states and encoding circuits, tailoring them to specific hardware noise models without relying on the stabilizer formalism. 
This data-driven framework provides rigorous accuracy bounds and enables the discovery of noise-adapted codes for NISQ devices that can outperform standard constructions, making it more suited for exploring new code structures while ensuring the learned codes satisfy fundamental QECC conditions.




\subsection{Early Fault-Tolerant Quantum Computing with Partial Quantum Error Correction (pQEC)}
As quantum hardware development transitions toward the EFT era, the distinction between error mitigation and correction is diminishing, motivating a shift toward hybrid frameworks.
Instead of treating them as disjoint techniques, recent strategies apply QEM to coarse-grained logical qubits protected by QECCs to clean up residual errors, thereby relaxing the stringent threshold requirements for ultimate fault tolerance~\cite{cai2023quantum, suzuki2022quantum}. 
This trend is evident in recent studies: state-preparation correction has been introduced to suppress coherent errors during optimization~\cite{wang2022state}, and the potential of QEM to enhance trainability by alleviating noise-induced BPs has been investigated~\cite{wang2024can}. 
Furthermore, circuit-division strategies have been proposed to co-optimize EM and partial QEC under constrained logical budgets~\cite{dutkiewicz2025error}. These developments suggest that EFT algorithms will rely on the co-design of mitigation and correction, serving as the necessary bridge to FT regime.

pQEC has emerged as a crucial intermediate step in the transition. 
Unlike the fault-tolerant QEC, which encodes the entire quantum algorithm in logical qubits (requiring significant overhead), pQEC focuses on encoding only the most noise-sensitive parts of the algorithm, such as specific gates or qubits, while using QEM to mitigate the impact of noise on the rest of the algorithm.
This allows for a more resource-efficient approach while still providing substantial improvements in error resilience.

Recent studies highlight the significant potential of pQEC to enhance the performance of VQAs despite the limitations of current hardware. 
In the partially fault-tolerant architecture~\cite{akahoshi2024partially}, only a subset of operations, namely Clifford gates, are fully protected using logical qubits. Arbitrary-angle (non-Clifford) rotations are implemented without magic-state distillation by teleporting a specially injected ancilla state via a repeat-until-success protocol.
Error mitigation is then applied to suppress the dominant rotation-induced noise.
Based on numerical simulations, researchers estimated that it is possible to perform roughly $1.72\times10^7$ Clifford operations and $3.75\times10^4$ arbitrary rotations with 64 logical qubits on early-FT devices consisting of $10^4$ physical qubits with error rate $p=10^{-4}$.
These numerical estimates suggest the potential for quantum advantage over classical methods under the assumed resource model.
Overall, this approach is practical and resource-efficient for near-term devices, improving error resilience without the overhead of fault-tolerant encoding throughout the computation.

VQAs have been implemented within this pQEC framework~\cite{dangwal2025variational}.
Their simulations demonstrated that this hybrid approach could improve the fidelity of VQE circuits by a factor of 9.27 relative to the version without the pQEC framework.
In addition, they propose architectural optimizations that reduce circuit latency by $\approx 2$x, and achieve qubit packing efficiency of $66\%$.

Furthermore, the pQEC framework has been applied to reduce the overhead of QEM and probabilistic error cancellation~\cite{dalfavero2025error}.
They showed that by adding the logical ancilla qubits, protected by QECC, into the quantum circuit, one can arbitrarily reduce the sampling overhead of probabilistic error cancellation in a continuous space-time tradeoff, even reaching \(O(1)\) samples, but only at the cost of exponential circuit depth. 
But they conjectured that any mitigation protocol with (sub-)polynomial sampling complexity must incur exponential time and/or space.
They also show an application to open-system simulation with about an order-of-magnitude fewer gates than current state-of-the-art methods.
Partial or hybrid error mitigation-correction has been applied to machine learning tasks. 
In quantum machine learning, researchers introduced a lightweight error detection method with a [[4,2,2]] code within a variational quantum classifier~\cite{adermann2026variational}, demonstrating improved learning accuracy under noisy conditions. While the method focuses on detection rather than full correction, it highlights the stabilizing effect of selective logical encoding in variational optimization.

In parallel, high-performance logical compilers have been developed to enable efficient mapping and scheduling of variational circuits onto large-scale logical qubit architectures~\cite{watkins2024high}. 
These advances indicate a trend toward architecture-aware variational design, where algorithmic expressivity and logical resource efficiency are jointly optimized.

Another emerging direction involves making error correction itself variational and trainable. Recent work demonstrates that both encoding and recovery maps can be optimized via a differentiable objective that maximizes state distinguishability, effectively treating QEC as a learnable component within hybrid workflows~\cite{meyer2025learning}. 
By integrating the adaptability of VQAs with the robustness of QEC, this approach represents a convergence of algorithmic optimization and fault-tolerance engineering. 
In the long term, such variational QEC frameworks may enable adaptive protection strategies tailored to specific noise channels or algorithmic requirements.

Although error rates are expected to decrease in the EFT era, the trainability of large-scale variational circuits remains a major challenge. 
Analyses of noise-induced and topology-dependent BPs reveal that these phenomena persist even under partial fault tolerance, imposing fundamental scaling limits on gradient-based optimization~\cite{singkanipa2025beyond}. 
Together, these studies show that, even without full encoding, partial QEC can significantly improve the stability and scalability of VQAs, allowing them to tackle more complex problems on today’s noisy near-term quantum devices.


\subsection{Fault-tolerant regime}

In the fault-tolerant regime, the role of VQAs changes substantially. Rather than serving primarily as a noise-resilient heuristic for NISQ devices, VQAs can become a high-accuracy algorithmic framework built on logical qubits protected by QEC. With greatly suppressed physical errors and access to deeper circuits, fault-tolerant VQAs may enable more reliable and expressive variational protocols for applications such as quantum chemistry, combinatorial optimization, and quantum machine learning.
In particular, fault-tolerant VQAs may serve as flexible logical-level solvers when exact quantum algorithms are too resource-intensive, offering a tunable trade-off between circuit expressibility and fault-tolerant cost.

In this setting, an important research direction is how to adapt variational algorithms to the resource constraints of fault-tolerant architectures. For example, fault-tolerant VQE schemes that explicitly minimize \(T\)-depth and \(T\)-count have been proposed, showing that resource-aware circuit construction can substantially reduce the overhead of logical operations while preserving algorithmic accuracy~\cite{sayginel2023fault}. More broadly, fault-tolerant implementations may allow VQAs to exploit longer coherent evolutions and more expressive ansatze than are feasible in the NISQ regime.

At the same time, fault tolerance does not automatically make VQAs efficient. A central challenge is the large logical-resource overhead required to compile variational circuits into fault-tolerant gate sets. For example, resource estimates for combinatorial optimization heuristics have shown that algorithms such as QAOA may require substantial logical overhead before achieving practical advantage~\cite{sanders2020compilation}. In addition, scalable fault-tolerant VQAs will require efficient methods for logical state preparation, circuit encoding and decoding, and measurement of cost functions within a QEC-protected architecture. Another open issue is whether the classical optimization loop can remain efficient when each function and gradient evaluation carries a much higher logical cost.

Overall, fault-tolerant quantum computing could significantly expand the scope and accuracy of VQAs, but their practical usefulness in this regime will depend on careful co-design of ansatz, compilation strategies, and optimization methods under realistic fault-tolerant resource constraints.

\section{Applications of VQAs}
\label{sec:applicationofVQAs}
Representative applications of VQAs can be grouped into physics, mathematical computing, chemistry, and information science, each with distinct cost functions, ansatz requirements, and hardware bottlenecks.

\begin{figure*}
    \centering
    \includegraphics[width=0.9\linewidth]{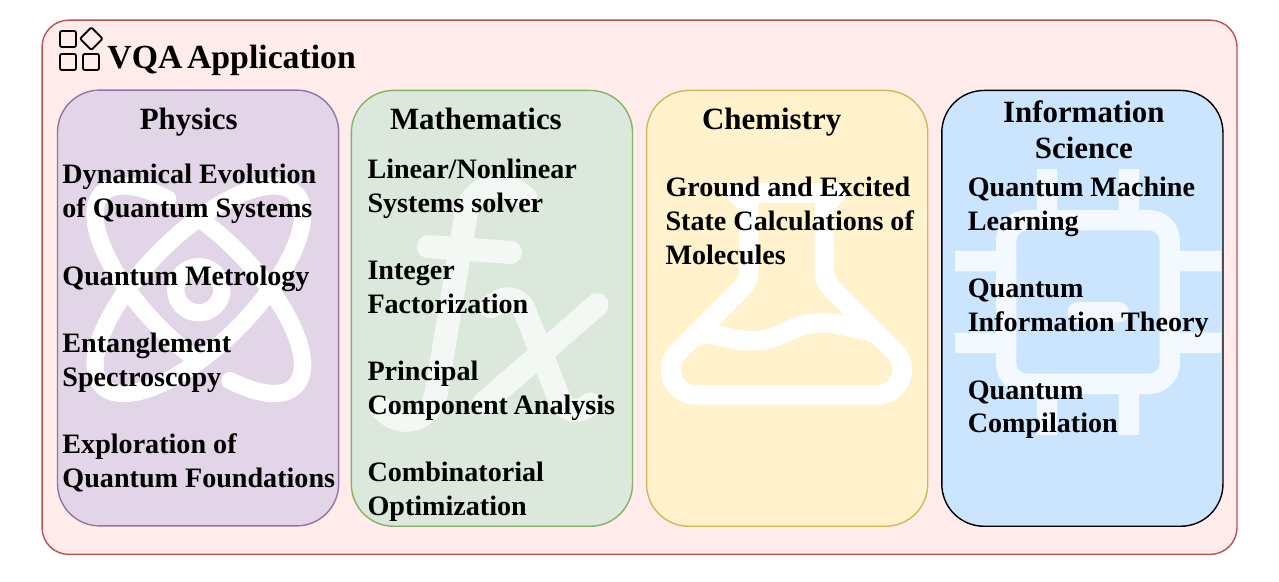}
    \caption{Applications of VQAs across major disciplines.}
    \label{VQAAPP}
\end{figure*}

\subsection{VQAs for physics}
VQAs have become an important framework for physics-oriented applications on near-term quantum devices. Owing to their hybrid quantum-classical nature, they are well suited to settings where circuit depth is limited but physically meaningful approximations remain accessible. In this context, VQAs have been applied to the simulation of dynamical evolution in quantum systems, to quantum metrology through the optimization of sensing protocols, to entanglement spectroscopy for characterizing many-body correlations, and to questions in quantum foundations, including the study of nonclassical features of quantum states and processes. Together, these directions illustrate the breadth of VQA-based methods as tools for exploring both practical and fundamental problems in physics.
\subsubsection{Dynamical evolution of quantum systems}
Simulations of dynamical evolution are among the most promising applications of quantum computers. VQAs have been widely used for real-time Schrödinger evolution of closed systems ~\cite{li2017efficient,cirstoiu2020variational,heya2023subspace}, imaginary time evolution ~\cite{mcardle2019variational}, and Lindblad dynamics of open systems ~\cite{watad2024variational,luo2024variational}.

Ying Li and Simon C Benjamin~\cite{li2017efficient} proposed an iterative variational method designed to simulate real-time quantum dynamics on near-term noisy quantum devices. This method represents the evolving state with a parameterized ansatz and uses the time-dependent variational principle to transform Schrödinger’s equation into differential equations for those parameters. Numerical simulations on the Ising model showed superior performance over standard Trotter methods under realistic noise. A variational fast forwarding technique~\cite{cirstoiu2020variational} was introduced to achieve fast forwarding of long-time simulations by variationally diagonalizing a short-time simulated unitary operator. The Subspace Variational Quantum Simulator~\cite{heya2023subspace} restricts simulation to the low-energy subspace via trained variational mapping and phase rotations, offering NISQ-friendly efficiency. For imaginary time, Sam McArdle \emph{et al.}~\cite{mcardle2019variational} used hybrid variational optimization with shallow circuits to simulate non-unitary evolution and effectively finds ground-state energies of molecular systems. Open-system approaches include a quantum superstate formalism~\cite{watad2024variational}, which encodes density matrices into larger pure states to facilitate processing within quantum circuits and a quantum state diffusion method~\cite{luo2024variational}, which stochastically unravels the dynamics and reduces quantum resource requirements.

The theoretical framework for variational quantum simulation has been extended to mixed states and both real and imaginary time evolution~\cite{yuan2019theory}. A neural network-based algorithm for simulating time evolution of closed and open quantum systems has been proposed~\cite{lee2021neural}. A unified framework for Variational Quantum Simulation of General Processes~\cite{endo2020variational} shows that many tasks, including real/imaginary time dynamics, non-Hermitian evolution, linear algebra problems, and open-system dynamics, can be framed as first-order generalized time-evolution equations. VQA performance in simulating spin-boson dynamics has been compared to the Trotter method~\cite{miessen2021quantum}, revealing that while VQAs are suitable for small systems due to linear growth in parameters and circuit depth, their scalability is limited by the increase in required measurements.

These variational methods have demonstrated practical advantage on NISQ hardware, especially for classical partial differential equations discretized as Hamiltonians. The advection-diffusion equation was simulated on IBM Fez superconducting processor using Trotterization, variational quantum time evolution (VarQTE), and adaptive variational quantum dynamics simulation (AVQDS)~\cite{alipanah2025quantum}, where VarQTE and AVQDS succeeded with much lower gate counts and circuit depth than Trotterization (which was infeasible due to excessive operations), and AVQDS used the fewest resources overall.

\subsubsection{Quantum metrology}
Quantum metrology seeks to exploit quantum resources such as entanglement and coherence to enhance parameter-estimation precision beyond classical limits. The ultimate precision bounds are commonly expressed in terms of the quantum Fisher information (QFI) and related quantities, providing a rigorous framework for both idealized and noisy sensing scenarios~\cite{giovannetti2011advances}. In practice, identifying probe states and measurement strategies that saturate these bounds is intractable under realistic noise, finite resources, and multiparameter conditions~\cite{escher2011general}. Decoherence, restricted control, and non-commuting generators further constrain the applicability of analytically derived optimal protocols. These challenges make quantum metrology a natural application domain for VQAs, which reformulate the search for near-optimal sensing strategies as an optimization over PQCs. Recent work demonstrated that variational techniques efficiently explore classically inaccessible state spaces, yielding highly entangled probes that outperform conventional resources under realistic noise models~\cite{koczor2020variational}.

Methodologically, VQA-based metrology extends beyond probe optimization to include measurement operators and, in some cases, control protocols. Variational probe design has been applied to single- and multiparameter tasks, including vector-field sensing and constrained-control scenarios, where tailored entangled states approach hardware-limited optimal performance~\cite{koczor2020variational,kaubruegger2023optimal,yang2022variational}. Recent frameworks enable joint optimization of probes and measurement settings, critical in noisy multiparameter regimes where optimal measurements depend explicitly on the noise model~\cite{meyer2021variational}. These approaches employ cost functions derived from classical or quantum Fisher information, mean-squared error, or Bayesian risk, and utilize parameter-shift rules to perform gradient-based optimization directly on quantum hardware~\cite{cimini2024variational,meyer2021variational}. Complementary efforts focus on variational QFI estimation, providing efficient certification and optimization for mixed states and imperfect devices~\cite{vitale2024robust,beckey2022variational,van2021measurement}.

The practical impact of variational metrology is evidenced by experimental demonstrations across platforms. Programmable trapped-ion sensors implementing VQAs achieve near-fundamental limits and outperform conventional spin-squeezing under realistic noise and control constraints~\cite{marciniak2022optimal}. In photonic and continuous-variable systems, VQAs combined with optical nonlinearities and reconfigurable interferometers enable scalable multiparameter estimation with enhanced noise robustness~\cite{kaubruegger2019variational,cimini2024variational,munoz2024photonic}. These studies exemplify a quantum-aided design approach, where quantum devices learn near-optimal strategies adapted to their native interactions and noise, rather than relying on idealized theoretical constructs. Variational approaches are expected to play an increasing role in adaptive and multiparameter sensing, as well as in integrating quantum metrology into feedback-stabilized quantum technologies and NISQ-era processors.

\subsubsection{Entanglement Spectroscopy}

Entanglement spectroscopy extracts the eigenvalue spectrum of reduced density matrices, offering a clear view of many-body quantum states. It excels at revealing topological order through spectral degeneracies or gaps that entanglement entropy alone often misses. VQAs suit this task well on NISQ hardware, using shallow circuits and hybrid optimization. Early methods include the variational quantum state eigensolver (VQSE) and variational quantum state diagonalization (VQSD), which variationally diagonalize reduced density matrices as mixed states~\cite{cerezo2022variational,larose2019variational}. The Quantum Singular Value Decomposer (QSVD) learns Schmidt singular values directly from bipartite pure states, aligning naturally with entanglement spectra~\cite{bravo2020quantum,garcia2024finding}, entanglement-informed ansatz designs~\cite{joch2025entanglement}, and cascaded VQE variants~\cite{gunlycke2024cascaded}.

Recent improvements include QSVD-based approaches and their variational implementations
for learning Schmidt spectra~\cite{bravo2020quantum,garcia2024finding},
as well as prescreening and hybrid eigenvector-learning strategies for reduced operators~\cite{wakaura2021prescreening}.
The typical workflow uses one VQA (often VQE) to approximate the ground state and another to extract its entanglement spectrum, enabling scalable detection of correlations and topological signatures.

Hardware demonstrations are emerging. VQSD was run on Rigetti superconducting hardware to diagonalize small states and simulate entanglement spectroscopy for a 12-spin Heisenberg chain ground state~\cite{larose2019variational}, VQSE has been tested in NISQ settings for single-copy density matrix eigenvalue learning~\cite{cerezo2022variational}.

\subsubsection{Quantum foundations}
VQAs are emerging as practical tools to probe foundational aspects of quantum mechanics. Early hybrid variational schemes demonstrated that core objects of quantum foundations, such as the decoherence functional in the Consistent Histories formalism, can be represented and optimized variationally, enabling classically intractable analyses of consistency and decoherence~\cite{arrasmith2019variational}.  

Recent developments have expanded these ideas along two complementary directions. First, algorithmic innovations like Schrödinger–Heisenberg hybrid VQAs combine shallow hardware circuits with classically tractable “virtual” Heisenberg circuits, enhancing expressibility and allowing variational exploration of highly entangled, dynamically evolving, or long-range correlated states relevant to foundational studies~\cite{shang2023schrodinger}. Second, variational methods have been applied to questions at the interface of information scrambling, learnability, and unitary compilation, revealing fundamental limitations in learning highly scrambling dynamics and providing insights for problems such as information recovery from Hawking radiation~\cite{anschuetz2022quantum,garcia2023resource}. In addition, exploratory approaches such as variational rewinding, which combine hybrid variational circuits with anomaly-detection protocols, have been proposed to investigate time-series data and signals relevant to gravitational-wave detection~\cite{rodrigues2025quantum}. 

\subsection{VQAs for Mathematical Problems}

Mathematical applications represent a key frontier for VQAs, which have been proposed to tackle problems including solving linear systems of equations, matrix-vector multiplication, nonlinear equations, integer factorization, principal component analysis (PCA), and optimization. The core goal of these VQAs is to achieve heuristic scalings comparable to the provable performance of fault-tolerant quantum algorithms while remaining compatible with the constraints of the NISQ era. 

\subsubsection{Quantum linear systems problem}
For solving the quantum linear systems problem (QLSP), where the task is to prepare a normalized state $|x\rangle$ such that $A|x\rangle\propto|b\rangle$ for an $N\times N$ linear system $Ax=b$, classical algorithms scale polynomially in $N$, while the Harrow-Hassidim-Lloyd quantum algorithm~\cite{harrow2009quantum} offers logarithmic scaling in $N$ but is impractical for near-term implementation due to excessive circuit depth. Motivated by this, VQAs for QLSP~\cite{bravo2023variational,huang2019near} assume $A=\sum_{k}c_{k}A_{k}$, construct a Hamiltonian $H_{G}=A(1-|b\rangle\langle b|)A^{\dagger}$ whose ground state is the QLSP solution, and minimize the cost function $C(\theta)=\langle\psi(\theta)|H_{G}|\psi(\theta)\rangle$. Though this cost landscape may suffer from BPs~\cite{larocca2025barren}, the issue can be mitigated by local Hamiltonians with the same ground state or hybrid ansatz strategies~\cite{huang2019near}. Numerical benchmarks already point toward logarithmic scaling in system size, which could translate into significant speedup compared with classical methods, even after accounting for BPs and their mitigations~\cite{larocca2025barren,huang2019near}. This theoretical outlook has gained real-world traction on Rigetti's superconducting processors, where variational linear solvers handled systems up to $1024 \times 1024$ while preserving the desired scaling under noisy conditions~\cite{bravo2023variational}. 

Related to QLSP, matrix-vector multiplication aims to prepare $|x\rangle\propto A|b\rangle$; here, the Hamiltonian $H_{\text{M}}=1-A|b\rangle\langle b|A^{\dagger}/\|A|b\rangle\|^{2}$ has the desired $|x\rangle$ as its zero-energy ground state~\cite{xu2021variational}, and the fidelity of an approximate solution $|\psi(\theta^{*})\rangle$ to the exact state can be lower-bounded as $|\langle\psi(\theta^{*})|x\rangle|^{2}\geq 1-\langle\psi(\theta^{*})|H_{\text{M}}|\psi(\theta^{*})\rangle$, enabling verification via small cost function values. 

\subsubsection{Nonlinear equations solver}
For nonlinear equations, two primary VQA approaches have been developed, with the first approach undergoing significant extensions in recent studies. The first approach, initially targeting the time-independent nonlinear Schrödinger equation, discretizes space into a finite grid, computes nonlinear functions using multiple copies of variational quantum states, and minimizes a total energy-based cost function~\cite{lubasch2020variational}. Recent advancements have expanded this framework to address a broader range of nonlinear systems: it now supports first-order time-dependent nonlinear partial differential equations and high-dimensional problems (e.g., 2D nonlinear Black-Scholes equations) by decomposing complex nonlinear operators into implementable quantum gates (e.g., Adder operators for spatial derivatives and Diagonal operators for point-wise multiplication) and training intermediate quantum states to handle intricate nonlinear terms~\cite{sarma2024quantum}. A key optimization to mitigate the exponential precision requirement in cost function estimation has also been proposed, which integrates the entire time evolution into a single optimization via an additional time-dimensional qubit register~\cite{pool2024nonlinear}. The second approach employs basis functions as nonlinear feature maps, prepares a quantum state representing linear combinations of these basis functions using a parameterized ansatz, computes function derivatives through the parameter-shift rule, and optimizes a cost function that enforces the nonlinear differential equation to hold at selected collocation points~\cite{kyriienko2021solving}. 

\subsubsection{Integer factorization}
For integer factorization, large-scale Shor's algorithm is near-term infeasible, so a VQA based on the QAOA was proposed, formulating factorization as an Ising model ground-state problem with numerical heuristics indicating that a linear number of ansatz layers ($p\in \mathcal{O}(n)$) yields large overlap with the ground state~\cite{anschuetz2019variational,anschuetz2022quantum}. Experiments on IBM superconducting devices have factored small integers and provided clear data on how noise affects circuit depth and resource demands~\cite{karamlou2021analyzing}.

\subsubsection{PCA}
In data science, PCA has also seen VQA advances: early quantum PCA algorithms~\cite{lloyd2014quantum} relied on quantum phase estimation and density matrix exponentiation, while subsequent variational approaches~\cite{larose2019variational,xin2021experimental} include a quantum state diagonalization algorithm (using a cost function quantifying Hilbert-Schmidt distance between $\hat{\rho}(\theta) = U(\theta)\rho U(\theta)^\dagger$ and the dephasing channel $\mathcal{Z}(\hat{\rho}(\theta))$, requiring $2n$ qubits) and a more qubit-efficient VQA (using $n$ qubits, exploiting the link between diagonalization and majorization to define $C(\theta) = \text{Tr}[\hat{\rho}(\theta)H]$ for a non-degenerate Hamiltonian $H$, where Schur concavity ensures minimization corresponds to diagonalization of $\hat{\rho}(\theta)$). Hardware demonstrations, including resonant qPCA on nitrogen-vacancy centers, have extracted principal components from small density matrices with high fidelity and remarkably few extra qubits~\cite{li2021resonant}.

\subsubsection{QAOA}

The QAOA, proposed by Farhi et al.~\cite{farhi2014quantum}, approximates solutions to combinatorial optimization problems (e.g., Max-Cut, SAT) by discretizing adiabatic evolution into alternating phase and mixing layers. Variational parameters are optimized classically to minimize the cost Hamiltonian expectation value, guiding the system toward the ground state encoding the optimal solution.

To adapt to NISQ constraints, enhancements improve expressibility and convergence. Multi-Angle QAOA (ma-QAOA) assigns unique parameters per term for better solution-space exploration and higher approximation ratios at shallow depths~\cite{herrman2022multi,gaidai2024performance}. QAOA+ adds problem-independent parameters to match deeper-circuit performance without extra noise~\cite{chalupnik2022augmenting}. Digitized Counterdiabatic QAOA (DC-QAOA) suppresses non-adiabatic transitions, reducing required layers; it has shown advantages over standard QAOA in molecular docking~\cite{chandarana2022digitized,ding2024molecular}. Adaptive Bias QAOA (ab-QAOA) uses learnable local fields in mixers to accelerate convergence~\cite{yu2022quantum}.

Dynamic variants include Adaptive Derivative-Assembled ADAPT-QAOA, which iteratively selects operators from a pool via gradients to minimize gates~\cite{zhu2022adaptive}, and constraint-preserving approaches like the Quantum Alternating Operator Ansatz (QAOAnsatz) with Grover-like mixers for feasible subspaces~\cite{hadfield2019quantum,bartschi2020grover}. Recursive QAOA (RQAOA) iteratively reduces problem size by measuring correlations~\cite{bravyi2020obstacles}. Constraint-preserving mixing Hamiltonians keep evolution within feasible subspaces~\cite{hao2026constraint}.

Initialization and optimization refinements mitigate BPs: Warm-Starting QAOA uses classical Semidefinite Programming (SDP) relaxations for parameters that meet or exceed classical guarantees, outperforming standard QAOA at low depths on IBM and Quantinuum hardware~\cite{egger2021warm}. Feedback-based Algorithm for Quantum Optimization (FALQON) employs Lyapunov control for monotonic improvement~\cite{magann2022lyapunov}. Spanning Tree QAOA and Quantum Dropout further refine the ansatz design and landscape~\cite{wurtz2021classically,wang2022quantum}.

These variants, validated on NISQ devices including IBM~\cite{saavedra2025quantum,cadavid2025bias}, Quantinuum trapped-ion~\cite{he2025performance}, and large-scale 127-qubit runs~\cite{pelofske2023quantum}, demonstrate robust performance despite noise, highlighting QAOA's evolving practicality for combinatorial tasks.



\subsection{VQAs for chemistry}
Because standard VQEs are intrinsically designed for ground-state optimization, a variety of extensions have been proposed to access excited states while remaining compatible with NISQ hardware. One prominent approach is folded-spectrum VQE (FS-VQE), which targets eigenstates within a prescribed energy window by minimizing a shifted squared Hamiltonian. This formulation allows excited states to be prepared using the ground-state ansatz and benefits from reduced measurement overhead through commuting Pauli grouping~\cite{cadi2024folded}. Numerical benchmarks on $\mathrm{H}_2$ and $\mathrm{LiH}$ have achieved chemical accuracy for multiple excited states, with early hardware demonstrations supporting its near-term feasibility. In addition, deflation-based methods provide an alternative route by adapting classical orthogonalization techniques to construct multiple excited states efficiently. Variational quantum packaged deflation reduces circuit depth and measurement costs while preserving orthogonality among states, whereas the orthogonal state reduction VQE enforces explicit constraints to prevent degeneracy and state overlap~\cite{xie2022orthogonal, wen2021variational}. Additionally, adaptive strategies further enhance excited-state capabilities by constructing problem-specific ansatzes on the fly. In particular, the adaptive VQE for highly excited states minimizes the energy variance, which vanishes only for exact eigenstates, rather than minimizing the energy itself. This strategy enables the direct targeting of arbitrary excited states without introducing additional classical penalty terms~\cite{zhang2021adaptive}. Although successful on both integrable and non-integrable Ising models, its scalability depends critically on the choice of operator pool and may require exponentially many parameters for highly excited states. Furthermore, complementary developments include full circuit-based eigenvalue solvers~\cite{wen2024full}, subspace-search quantum imaginary time evolution methods that propagate orthogonal subspaces~\cite{cianci2024subspace}, feedback-based optimization schemes that leverage measurement outcomes~\cite{rahman2024feedback}, and measurement-optimization techniques that further alleviate sampling overhead~\cite{choi2023measurement}. Collectively, these advances indicate that folded-spectrum, deflation, adaptive, and equation-of-motion-based methods are rapidly maturing, making excited-state quantum chemistry increasingly feasible on NISQ hardware despite noise and measurement constraints.

Recent experimental demonstrations on NISQ platforms further substantiate this progress. On superconducting devices, spin-restricted ansatzes with automatically adjusted constraints have enabled molecular excited-state calculations with improved symmetry preservation and reduced state mixing~\cite{gocho2023excited}. The contextual subspace VQE has produced accurate dissociation curves, including excited states of molecular nitrogen, on IBM superconducting processors~\cite{weaving2025contextual}. Beyond superconducting architectures, the photonic VQE has achieved high-fidelity ground and excited state calculations~\cite{hu2025photonic}, while Variational Quantum Eigensolver-Equation of Motion (VQE-qEOM) has been used to compute vertical singlet excited states of battery electrolytes ($\mathrm{LiPF}_6$, $\mathrm{NaPF}_6$) on near-term quantum hardware~\cite{hossain2025quantum}. Together, these results confirm the practical viability of the excited-state VQE variants on noisy quantum devices.
\subsection{VQAs for information science}
\subsubsection{QML}
Quantum Machine Learning (QML) integrates quantum computing with machine learning, utilizing VQAs to potentially achieve computational advantages. This section reviews VQA-based models across supervised, unsupervised, semi-supervised, and reinforcement learning frameworks.

\paragraph{Supervised Learning.}
Supervised VQC-based learning has evolved from foundational Variational Quantum Models (VQM) and Quantum Circuit Learning (QCL)~\cite{mitarai2018quantum}, which demonstrated early success in binary classification~\cite{farhi2018classification}, to architectures addressing NISQ constraints via dimensionality reduction~\cite{chen2020hybrid}. Structural advancements include hierarchical classifiers like Quantum Tree Tensor Networks (TTN) and Multi-scale Entanglement Renormalization Ansatz (MERA)~\cite{grant2018hierarchical}. These were further optimized by Chen et al.~\cite{chen2021end} using MPS for feature extraction and Qi et al.~\cite{qi2023qtn} via Tensor Product Encoding. Hybrid approaches, such as the Variational Quantum Tensor Networks (VQTN) algorithm, integrate classical post-processing to reduce complexity~\cite{huang2021variational}. Regularization techniques, including structural constraints and Dropout, have also been adapted for quantum circuits~\cite{schuld2020circuit, chen2020hybrid}.

Quantum Convolutional Neural Networks represent a critical direction due to their immunity to BPs ~\cite{pesah2021absence}. While foundational architectures were established by Cong et al.~\cite{cong2019quantum} and Oh et al.~\cite{oh2020tutorial}, and optimized by Wei et al.~\cite{wei2022quantum}, practical implementation often relies on hybrid quantum-classical frameworks~\cite{liu2021hybrid, chen2023quantum}. Other innovations include Variational Shadow Quantum Learning (VSQL) for efficient feature extraction~\cite{li2021vsql} and quantum analogues of perceptrons~\cite{monteiro2021quantum, chu2022qmlp} and graph networks~\cite{hu2022design}. Notably, Hybrid Neural Networks (HNN) treating VQCs as individual neurons have shown improved simulation accuracy~\cite{arthur2022hybrid}.

\paragraph{Unsupervised Learning.}
Unsupervised VQAs focus on generative modeling and compression. The Quantum Circuit Born Machine (QCBM) models data distributions using the probabilistic nature of wave functions~\cite{benedetti2019generative}. Improvements using differentiable Maximum Mean Discrepancy (MMD) loss~\cite{liu2018differentiable} have enabled applications in finance~\cite{gong2022born, coyle2021quantum, vcepaite2022continuous, alcazar2020classical, kondratyev2021non}. The Variational Quantum Boltzmann Machine (VarQBM) approximates Gibbs states to simulate thermal distributions~\cite{zoufal2021variational, shingu2021boltzmann}. Meanwhile, Quantum Autoencoders (QAEs) facilitate dimensionality reduction~\cite{romero2017quantum, huang2020realization}. Recent work addresses trainability via local cost functions~\cite{cerezo2021cost, mcclean2018barren} and introduces efficient encoding architectures for drug discovery~\cite{li2022scalable} and data approximation~\cite{bravo2021quantum, wu2023efficient}.

\paragraph{Semi-supervised Learning.}
Quantum semi-supervised learning is primarily realized through Quantum Generative Adversarial Networks (QGANs), adapting the adversarial framework to VQCs~\cite{goodfellow2014generative, lloyd2018quantum}. Optimization involves a min-max game to minimize the trace distance between generated and real states~\cite{dallaire2018quantum, tian2023recent}. Fully quantum architectures offer potential exponential speedups for specific tasks~\cite{lloyd2018quantum}, with experimental proofs achieving high fidelity~\cite{hu2019quantum}. To enhance stability, researchers have incorporated Wasserstein metrics~\cite{chakrabarti2019quantum} and developed variants like PATCH-GAN~\cite{huang2021experimental}. Applications span discrete~\cite{romero2018strategies, zeng2019learning, situ2020quantum} and continuous variables~\cite{romero2021variational, li2021quantum}, with methods to optimize data loading efficiency~\cite{zoufal2019quantum}. Newer models like Enhanced Quantum Generative Adversarial Network ~\cite{niu2022entangling}, Quantum Molecule Generative Adversarial Network~\cite{kao2023exploring}, and hybrid anomaly detection systems~\cite{herr2021anomaly, tsang2023hybrid, du2019efficient} continue to expand QGAN utility on NISQ devices.

\paragraph{Reinforcement Learning.}
Quantum Reinforcement Learning (QRL) utilizes VQCs for function approximation. Value-based methods, such as Variational Quantum Deep Q-learning (VQ-DQN)~\cite{chen2020variational} and Quantum Long Short-Term Memory Deep Recurrent Q-Network (QLSTM-DRQN)~\cite{chen2023quantum}, approximate Q-values to reduce parameter counts compared to classical benchmarks~\cite{mnih2013playing, van2016deep}. Techniques include transforming inputs into rotation angles~\cite{lockwood2021computational} and using trainable weights for value matching~\cite{skolik2022quantum}. Policy-gradient methods directly optimize actions~\cite{jerbi2021parametrized}, potentially offering faster parameter updates~\cite{jerbi2022quantum}, while Actor-Critic frameworks combine these approaches~\cite{wu2025quantum, kwak2021introduction}. Recent advances target robustness against noise~\cite{skolik2023robustness} and multi-agent systems, where centralized training strategies have demonstrated performance gains over classical counterparts~\cite{yun2022quantum, yun2023quantum}.

\subsubsection{Quantum information theory}

VQAs have emerged as a unifying framework in quantum information theory (QIT) for characterizing diverse quantum resources, particularly in the NISQ era. By employing shallow parameterized circuits and hybrid quantum–classical optimization, these methods recast resource quantification, often involving extremal decompositions, convex-roof constructions, or spectral estimation, into experimentally feasible optimization tasks. Uhlmann-theorem-inspired VQAs enable the estimation of Bures entanglement and reconstruction of closest free states while naturally extending to broader resource theories including quantum coherence, total correlations, and quantum discord~\cite{friedrich2025variational}. Variational trace distance estimation and variational fidelity estimation~\cite{chen2021variational,cerezo2020variational} offer scalable access to similarity measures that underpin quantum benchmarking, hypothesis testing, and metrology. More general-purpose VQAs, such as variational quantum thermalizers for learning modular Hamiltonians~\cite{verdon2019quantum} and QAEs for quantum data compression~\cite{romero2017quantum}, demonstrate that tasks like entropy estimation, state decorrelation, and structural learning can also be addressed within a variational framework, with QAEs additionally capturing aspects of channel encoding.

A major class of applications concerns quantum entanglement, where VQAs provide practical detection and quantification methods beyond the capabilities of conventional tomography. Variational entanglement detection, variational logarithmic-negativity estimation, and related hybrid approaches~\cite{chen2023near,wang2022detecting} enable efficient analysis of both pure and mixed states. Additional strategies include adversarial bipartite detection protocols~\cite{yin2022efficient}, steering-based variational quantum separability algorithms~\cite{philip2024schrodinger}, and Hilbert–Schmidt-distance-based variational separability verifiers~\cite{consiglio2022variational}. For multipartite systems, variational estimators of geometric entanglement~\cite{munoz2022variational} provide a scalable approach to measuring multiqubit entanglement, while BP-resistant local-to-global optimization strategies~\cite{zambrano2024avoiding} extend this framework and avoid BPs, enabling robust performance in large, noisy systems. Convex-roof–motivated variational schemes~\cite{androulakis2022variational} illustrate how sampling over parameterized ensembles can approximate mixed-state entanglement measures, while interference-enhanced variational protocols~\cite{zhang2025entanglement} show improved detection power under realistic experimental noise.

Taken together, these developments point toward a coherent paradigm in which VQAs serve as versatile tools for probing quantum properties that are classically intractable yet operationally essential for quantum information science. Integrating techniques such as local cost functions, quasi-probability decompositions, mid-circuit measurement, adversarial learning, and quantum machine-learning architectures~\cite{scala2022quantum}, VQAs have demonstrated broad experimental viability on superconducting devices~\cite{chen2023near}, IBM quantum hardware~\cite{munoz2022variational,zambrano2024avoiding}, and photonic platforms~\cite{yin2022efficient,zhang2025entanglement}. As the field continues to advance, variational approaches are poised to play a central role in quantifying quantum resources, validating near-term quantum processors, and enabling scalable protocols for quantum state characterization and quantum data learning.

\subsubsection{Quantum compilation}
In quantum compilation, the goal is to approximate a target unitary \(V\) using a native gate sequence \(U(\boldsymbol{\theta})\) with minimal or near-optimal circuit depth. This is especially important in the NISQ setting, where errors typically accumulate with circuit depth and two-qubit gate count under realistic noise models~\cite{bharti2022noisy,cai2023quantum}.

Quantum compilation tasks can be broadly categorized into full unitary matrix compiling (FUMC) and fixed input state compiling (FISC). FUMC aims to approximate the entire target unitary \(U \in \mathrm{SU}(2^n)\) by minimizing process-based distances, such as \(1 - |\mathrm{Tr}(U^\dagger V(\boldsymbol{\theta}))|^2 / d^2\), thereby ensuring correctness for arbitrary input states. In contrast, FISC focuses on a fixed input state \(\ket{\psi_{\mathrm{in}}}\) or a relevant subset of input states, and optimizes fidelity only within the corresponding subspace.

Within both settings, variational quantum algorithms (VQAs) provide a natural hybrid quantum-classical framework for optimization on NISQ hardware. In this context, Khatri et al.\ proposed quantum-assisted quantum compiling for tasks including circuit-depth compression, uploading unknown unitaries, noise tailoring, and benchmarking~\cite{khatri2019quantum}.

For FUMC, variational circuits can approximate arbitrary unitaries, as demonstrated on IBM superconducting processors where VQA-based compiled gates achieved improved process fidelity for noisy two-qubit operations~\cite{sharma2020noise}. However, scalability remains a major challenge. Madden showed that a sketch-and-solve algorithm significantly outperforms standard optimization methods in terms of circuit width and depth~\cite{madden2022sketching}. Local variational quantum compilation (LVQC) further enables large-scale time-evolution compilation through optimization on smaller subsystems, and naturally belongs to FUMC since it is designed to work for arbitrary input states~\cite{mizuta2022local}.

FISC-based methods reduce compilation overhead by restricting the objective to a relevant subspace. Experiments on Rigetti hardware demonstrated circuit-depth reductions for state-specific compilation tasks~\cite{khatri2019quantum}. Bilek et al.\ further proposed recursive variational quantum compiling (RVQC) for deep noisy circuits, where standard VQC may fail~\cite{bilek2022recursive}. Although primarily formulated for FISC, RVQC can be extended to FUMC by adopting an appropriate cost function.

Overall, variational quantum compilation provides a flexible framework for both full-unitary and task-specific compilation. Future work may focus on hardware-aware ansatz design, pulse-level compilation, and standardized cross-platform frameworks.

\section{Conclusion}
VQAs have emerged as one of the most versatile and promising paradigms for near-term quantum computing, effectively bridging the gap between NISQ devices and practical quantum advantage. Recent years have witnessed significant progress in algorithmic design, including the development of more expressive ansatzes, noise-resilient optimization strategies, and hybrid quantum-classical frameworks that enhance scalability and performance. Techniques such as adaptive ansatzes, gradient-based and gradient-free optimization, and error mitigation have demonstrated that VQAs can tackle problems in quantum chemistry, combinatorial optimization, and machine learning with unprecedented flexibility.
Looking forward to the FT era, VQAs are expected to retain a strategic role. Even in the presence of large-scale, error-corrected qubits, the variational approach offers unique advantages: it can serve as a resource-efficient tool for pre-processing, state preparation, and circuit compilation, while also enabling exploration of complex problem landscapes that would otherwise be intractable. Moreover, hybrid variational methods could be seamlessly integrated with fault-tolerant algorithms to accelerate convergence, reduce circuit depth, and mitigate computational overhead. Overall, the trajectory of VQAs suggests a continuum from NISQ to FTQC applications, where their adaptability and hybrid nature will remain central to harnessing the full power of quantum computation.

\section*{Author Contributions}
Z.W., J.H., and R.Y. contributed equally to this work and share first authorship. X.Y., and Y.Y. were involved in the conception and design of the study. 
Y.H., J.H., Z.W., and J.G. contributed to the writing of the overview of the VQA architecture. 
Y.Y., J.H., Y.Z., and Q.L. contributed to the discussion of the ultimate FT regime with VQAs. 
The applications of VQAs were written collaboratively by Z.W., J.H., R.Y., Q.D., Y.H., T.Z., Y.Z., and Y.Y., with each author contributing to different parts. 
T.Z. was responsible for collecting and organizing the literature related to the experimental studies. 
Z.W. and Y.Y. designed and prepared all figures in the manuscript. Y.Y., Q.D., and X.Y. contributed to the writing of the Introduction and Conclusion. All authors reviewed and approved the final manuscript.

\section*{Acknowledgments}
This work is supported by 
the National Natural Science Foundation of China (Grant Nos.~12504406, 12361161602, and 12342502), 
the NSAF (Grant No.~U2330201), 
the Beijing Natural Science Foundation (Grant Nos.~Z250004, 1254053, and 1264065), 
the Beijing Science and Technology Planning Project (Grant No.~Z25110100810000),
the Quantum Science and Technology-National Science and Technology Major Project (Grant No.~2023ZD0300200).

\bibliographystyle{unsrt}
\bibliography{ref}

\end{document}